\documentclass[onecolumn,floatfix,aps]{revtex4}
\usepackage{amssymb,amsmath}
\newcommand{\bea}{\begin{array}}
\newcommand{\ear}{\end{array}}
\newcommand{\bege}{\begin{equation}}
\newcommand{\enge}{\end{equation}}

\newcommand{\beq}{\begin{eqnarray}}\newcommand{\benu}{\begin{enumerate}}\newcommand{\enu}{\end{enumerate}}
\newcommand{\eeq}{\end{eqnarray}}

\newtheorem{remark}{Remark}

\begin{document}

\title{Diffeomorphism Invariance and Local Lorentz
Invariance}

\author{Rold\~ao da Rocha}
\email{roldao@ifi.unicamp.br}
\affiliation{IFGW, Universidade Estadual de Campinas,\\
CP 6165, 13083-970 Campinas, SP, Brazil.}

\author{Waldyr A. Rodrigues, Jr.}
\email{walrod@mpc.com.br,walrod@ime.unicamp.br}
\affiliation{Departamento de Matem\'atica Aplicada, IMECC\\
Universidade Estadual de Campinas,\\
CP 6065, 13083-859, Campinas, SP, Brazil.}


\begin{abstract}
We show that diffeomorphism invariance of the Maxwell and the Dirac-Hestenes
equations implies the equivalence among different universe models such that if
one has a linear connection with non-null torsion and/or curvature the others
have also. On the other hand \textit{local} Lorentz invariance implies the
surprising equivalence among different universe models that have in general
different $G$-connections with different curvature and torsion tensors.
 
\end{abstract}

\maketitle

\section{Introduction}

In this paper, by using the Clifford and spin-Clifford bundle formalism we
present a thoughtful analysis on the concepts concerning diffeomorphism
invariance and local Lorentz invariance of Maxwell and Dirac-Hestenes
equations. Diffeomorphism invariance implies the equivalence among different
universe models such that if one has non-null torsion and curvature, the
others also possess similar characters. Local Lorentz invariance implies the
astounding equivalence between different universe models that have in general
different $G$-connections with different curvature and torsion tensors. This
article is organized as follows: after presenting some algebraic
preliminaries in Section 2, in Section 3 the invariance of the Maxwell
Lagrangian and of Dirac-Hestenes equation, under diffeomorphisms, is
investigated from the extensor field formalism viewpoint. Lorentz
transformations and the Lienard-Wiechert formul\ae \thinspace\ are derived
in this context. In Section 4 active local Lorentz mappings are introduced,
regarding their action on electromagnetic fields. The covariant derivative
acting on vector and spinor fields is briefly revisited in the light of the
Clifford and spin-Clifford bundle context in Section 5. Next, using that
formalism we present in Section 6 the Dirac-Hestenes equation in
Riemann-Cartan spacetimes and recall that, in general, in theories of that
kind the spin generates torsion. Indeed, it is always emphasized that in
a theory where, besides the spinor field, also the tetrad fields and the
connection are dynamical variables, the torsion is not zero, because its
source is the spin associated with the spinor field. However, in Section 7 we show that
to suppose the Dirac-Hestenes Lagrangian is invariant under active
rotational gauge transformations implies in an equivalence between torsion
free and non-torsion free $G$-connections, and also that we may also have
equivalence between spacetimes with null and non-null curvatures.

\section{Some Preliminaries}

\label{sec2} A Riemann-Cartan spacetime is a pentuple $(M,\ $\textbf{g}$%
,\nabla,\tau_{\text{\textbf{g}}\ },\uparrow)$ where $(M,$\ \textbf{g}$%
\mathbf{)}$ is an oriented (by $\tau_{\text{\textbf{g}}}\in\sec
\bigwedge\nolimits^{4}T^{\ast}M$) and time-oriented (by the equivalence
relation $\uparrow$) 4-dimensional Lorentzian manifold $M,$ equipped with a
Lorentz metric \textbf{g }$\in\sec T_{2}^{0}M$ of signature $(1,3)$. The
operator $\nabla$ denotes the Levi-Civita (metric compatible) connection of 
\textbf{g} $(\nabla$\textbf{g}$=0),$ and in general $\mathbf{T}(\nabla)\neq
0 $, and $\mathbf{R}(\nabla)\neq0$, where $\mathbf{T}$ is the torsion tensor
of $\nabla$ and $\mathbf{R}$\textbf{\ }is the\textbf{\ }Riemann curvature
tensor of $\nabla$. When $\nabla$\textbf{g}$=0$ and $\mathbf{T}(\nabla)=0$
the pentuple $(M,\ $\textbf{g}$,\nabla,\tau_{\text{\textbf{g}}\ },\uparrow)$
is called a Lorentzian manifold. A Lorentzian manifold for which $M\simeq 
\mathbb{R}^{4}$ is called a Minkowski spacetime. In this case it is
represented by a pentuple $(M,\mbox{\boldmath{$\eta$}},\nabla,\tau _{%
\mbox{\boldmath{$\eta$}}},\uparrow)$\footnote{%
Another popular representation of Minkowski spacetime is the structure $(M,%
\mathbf{V,}$ $\mathbf{\bullet}$) where $\mathbf{M}=(M,\mathbf{V})$ is an
affine space and $\mathbf{V\simeq }\,\mathbb{R}^{4}$ is a vector space,
endowed with a Lorentzian scalar product $\bullet$ of signature $(1,3)$ and
which is oriented and time-oriented. This will be seen in the Appendix.}.
More details if needed can be found, e.g., in \cite{sawu}.

At each point $\mathfrak{e}\in M,$ we denote respectively by $T_{\mathfrak{e}%
}M$\textbf{\ }and $T_{\mathfrak{e}}^{\ast }M,$ the tangent and cotangent
spaces. Reference frames are time-like vector fields (pointing to the
future) in the world manifold $M.$ If $\mbox{\boldmath{$\eta$}}\in \sec
T_{2}^{0}M$ is the metric of Minkowski spacetime, there exists in $M$ a
global chart $(M,\varphi )$ with coordinate functions $\{x^{\mu }\}$ (said
to be in the \textit{Einstein-Lorentz-Poincar\'{e} gauge}) such that for the
section $\{e_{\mu }=\frac{\partial }{\partial x^{\mu }}\}$ of the
orthonormal frame bundle $\mathbf{P}_{\mathrm{SO}_{1,3}^{e}}(M,%
\mbox{\boldmath{$\eta$}})$ we have 
\begin{equation}
\mbox{\boldmath{$\eta$}}(e_{\mu },e_{\nu })=\mathrm{diag}(1,-1,-1,-1)
\label{0}
\end{equation}%
The pair $(T_{\mathfrak{e}}M\mathbf{,}\left. \mbox{\boldmath{$\eta$}}%
\right\vert _{\mathfrak{e}})\simeq \mathbb{R}^{1,3}$ is called Minkowski 
\textit{vector} space. The existence of global coordinates in the
Einstein-Lorentz-Poincar\'{e} gauge permits to identify all tangent (and
cotangent) spaces for all $\mathfrak{e}\in M$.\footnote{%
This will be used in the Appendix.}

Special Relativity (SR) refers to theories that have the Poincar\'{e} group
as a symmetry group\footnote{%
Details on the meaning of this statement can be found, e.g., in \cite{rosha}.%
}. This theory asserts that there is a class of \textit{physically equivalent%
} reference frames, the \textit{inertial} ones\footnote{%
The definition of reference frame in general, and inertial reference frame
in particular will be given below.}. The Clifford algebra associated with $%
\mathbb{R}^{1,3}$ is denoted by $\mathbb{R}_{1,3}\simeq\mathbb{H(}2\mathbb{)}
$ and is called the \textit{spacetime} algebra. The \textit{Dirac} algebra
is $\mathbb{C}\otimes\mathbb{R}_{1,3}\simeq\mathbb{R}_{4,1}\simeq\mathbb{C(}%
4)$, the Clifford algebra associated with a 5-dimensional vector space
endowed with a scalar product of signature $(4,1)$\footnote{%
For details, see, e.g., \cite{lounesto,rod2004}.}. Note that given a general
Riemann-Cartan spacetime we also have that $\mathcal{C}\ell ( T_{\mathfrak{e}%
}M,\left. \text{\textbf{g}}\right\vert _{\mathfrak{e}})=\mathbb{R}_{1,3}$.
Also, if \texttt{g}\textbf{\ }$\in\sec T_{2}^{0}M$ is the metric associated with the {%
cotangent} bundle, we have $\mathcal{C}\ell(T_{\mathfrak{e}}^{\ast}M,\left. 
\mathtt{g}\right\vert _{\mathfrak{e}})=\mathbb{R}_{1,3}$.

Fields in the Clifford algebra formalism\footnote{%
See, e.g., \cite{moro2004}, for a rigorous definition of Clifford and spinor
fields.} can be taken as sections of the Clifford bundle of multivectors,
denoted by $\mathcal{C}\ell (M,$\textbf{g}$)=\cup_{\mathfrak{e}}\mathcal{%
C\ell(}T_{\mathfrak{e}}M,\left. \text{\textbf{g}}\right\vert _{\mathfrak{e}%
}) $ or as sections of the Clifford bundle of multiforms, denoted by $%
\mathcal{C}\ell(M,\mathtt{g})=\cup _{\mathfrak{e}}\mathcal{C}\ell(T_{%
\mathfrak{e}}^{\ast}M,\left. \mathtt{g}\right\vert _{\mathfrak{e}}),$ which
we shall use in what follows, because it is more convenient for our
purposes. By $\mathcal{C\ell}^{0}\mathcal{(}M,\mathtt{g})$ we denote the
even subalgebra of $\mathcal{C}\ell(M,\mathtt{g})$.\footnote{%
Of course, the Clifford algebra of multiforms associated with a Minkowski
spacetime will be denoted by $\mathcal{C}\ell (M,\mathtt{\eta}),$ where $%
\mathtt{\eta}$ is the metric of the cotangent bundle.} Note that $\mathcal{%
C\ell}^{0}\mathcal{(}T_{\mathfrak{e}}^{\ast }M,\left. \mathtt{g}\right\vert
_{\mathfrak{e}})\simeq\mathbb{R}_{1,3}^{0}\simeq\mathbb{R}_{3,0}$, where $%
\mathbb{R}_{3,0}\simeq\mathbb{C(}2\mathbb{)}$ is the Pauli algebra. Then, a
Clifford field of multiforms will be considered as a section 
\begin{equation}
C\in\sec\bigwedge T^{\ast}M\hookrightarrow\sec\mathcal{C}\ell(M,\mathtt{g})
\label{1}
\end{equation}
where $\bigwedge T^{\ast}M={\displaystyle\oplus_{p=0}^{4}}%
\bigwedge^{p}T^{\ast}M$ denotes the exterior algebra of multiforms. The
symbol $\hookrightarrow$ means that $\bigwedge T^{\ast}M$ is {embedded} in $\mathcal{C}\ell(M,\mathtt{g})$.

A metric compatible connection $\nabla$ acting on the tensor bundle defines
a covariant derivative acting on Clifford fields, see e.g., \cite{crume}. A
reference frame\footnote{%
Please do not confuse this concept with the concept of a frame, that is a
section of the frame bundle.} $Z:M\supset U\rightarrow TM$ is a vector field
such that $\boldmath{\eta}(Z,Z)>0$. In Minkowski spacetime there are
infinite global inertial reference frames\footnote{%
Inertial reference frames did not exists in general in an arbitrary
Lorentzian spacetime, see, e.g., \cite{sawu,rool}.}. These are reference
frames for which $\nabla_{e_{\nu}}Z=0$, for $\nu=0,1,2,3$. The vector fields 
$\{e_{\mu}=\frac{\partial}{\partial x^{\mu}}\},$ with $\{x^{\mu }\} $ global
coordinates in the Einstein-Lorentz -Poincar\'{e} gauge satisfy $%
\nabla_{e_{\mu}}e_{\nu}=0$ and thus $e_{0}$ is qualified as an inertial
reference frame. The dual basis of $\{e_{\mu}\}$ will be denoted by $%
\{\gamma^{\mu}\}$. We assume that $\gamma^{\mu}\in\sec\bigwedge^{1}T^{\ast
}M\hookrightarrow\mathcal{C}\ell(M,\mathtt{\eta})$. The reciprocal basis related to $%
\{\gamma^{\mu}\}$ is denoted by $\{\gamma_{\mu}\}$ and is such that $\mathtt{%
\eta}(\gamma_{\mu},\gamma^{\nu}):=\gamma_{\mu}\cdot\gamma^{\nu
}=\delta_{\mu}^{\nu}$. Given a inertial frame, say $e_{0},$ it will be
represented in $\mathcal{C}\ell(M,\mathtt{\eta})$ by the physically
equivalent field $\gamma_{0}=\mbox{\boldmath{$\eta$}}(e_{0},\;)$ that by
abuse of language we also call a reference frame.

Let $\mathbf{L}$ be an arbitrary proper orthochronous Lorentz
transformation. A reference frame 
\begin{equation}
\gamma_{0}^{\prime}=\mathbf{L}\gamma_{0}=L_{0}^{\mu}\gamma_{\mu}  \label{2}
\end{equation}
is also an inertial reference frame. In the Clifford algebra formalism we
can write Eq.(\ref{2}) as 
\begin{equation}
\gamma_{0}^{\prime}=R\gamma_{0}\tilde{R},  \label{3}
\end{equation}
where $R\in\sec$\textrm{Spin}$_{3,1}^{e}(M)$, i.e., for any $\mathfrak{e}\in
M,$ $R(\mathfrak{e})\tilde{R}(\mathfrak{e})=1$, $R(\mathfrak{e})\in$ \textrm{%
Spin}$_{3,1}^{e}\simeq$ SL$(2,\mathbb{C)}$.

For a general Riemann-Cartan spacetime $(M,\ $\textbf{g}$,\nabla ,\tau_{%
\text{\textbf{g}}\ },\uparrow)$ we denote by $\{\mathbf{e}_{\mathbf{a}%
}\}\in\sec\mathbf{P}\mathrm{SO}_{1,3}^{e}(M)$ an orthonormal frame and by $%
\{\gamma^{\mathbf{a}}\}\in\sec P\mathrm{SO}_{1,3}^{e}(M)$ the respective
orthonormal coframe. The Dirac operator acting on sections of $\mathcal{C}\ell (M,\mathtt{g})$ is the invariant differential operator which maps
Clifford fields in Clifford fields, given by%
\begin{align}
{\mbox{\boldmath$\partial$}} & =\gamma^{\mathbf{a}}\nabla_{\mathbf{e}_{%
\mathbf{a}}},  \label{4} \\
{\mbox{\boldmath$\partial$}}C & ={\mbox{\boldmath$\partial$}}\mathbf{\wedge }%
C+{\mbox{\boldmath$\partial$}}\mathbf{\lrcorner}C.  \label{4bis}
\end{align}
When $\nabla$ is the Levi-Civita connection of \textbf{g}, we have 
\begin{equation}
{\mbox{\boldmath$\partial$}}=\mathbf{{\mbox{\boldmath$\partial$}}\wedge }+{%
\mbox{\boldmath$\partial$}}\mathbf{\lrcorner}=d-\text{ }\underset{\text{%
\textbf{g}}}{\delta},  \label{4biss}
\end{equation}
where $\underset{\text{\textbf{g}}}{\delta}$ is the Hodge coderivative
operator. Thus, in this case we can write%
\begin{equation}
{\mbox{\boldmath$\partial$}}C=dC-\text{ }\underset{\text{\textbf{g}}}{\delta 
}C.  \label{4bisss}
\end{equation}
Recall that coordinates functions for $U\subset M$ are mappings $x^{\mu
}:M\supset U\rightarrow\mathbb{R}$. These mappings can be considered as
sections, $x^{\mu}\in\sec\bigwedge^{0}T^{\ast}U\hookrightarrow\mathcal{C\ell
(}U,\mathtt{g})$. In the case of a Minkowski spacetime a special set of
coordinates naturally adapted to an inertial frame are the ones in the
Einstein-Lorentz-Poincar\'{e} gauge\footnote{%
See \cite{rosha} for details of the concept of coordinates naturally adapted
to a given general frame $Z.$}. They are global coordinate functions such
that 
\begin{equation}
{\mbox{\boldmath$\partial$}}x^{\mu}=\gamma^{\mu}.  \label{6}
\end{equation}
In this case the Dirac operator can be written as ${\mbox{\boldmath$%
\partial$}}=\gamma^{\mu}\nabla_{\mathbf{e}_{\mu}}=\gamma^{\mu}\partial_{%
\mu}. $

\section{Maxwell Theory and Diffeomorphism Invariance}

\label{sec3} Classical Maxwell theory on a Lorentzian spacetime deals with
an electromagnetic field $F\in\sec\bigwedge^{2}T^{\ast}M\hookrightarrow \sec
\mathcal{C}\ell(M,\mathtt{g})$ generated by a current $J\in\sec
\bigwedge^{1}T^{\ast}M\hookrightarrow\sec\mathcal{C}\ell(M,\mathtt{g}),$ and
the motion of \textit{probe} charges modelled by triples\footnote{%
The $m_{i}\in\mathbb{R}^{+}$ are the masses and the $q_{i}\in\mathbb{R}%
-\{0\} $ are the charges of the particles.} $(m_{i},q_{i},\sigma_{i})$ in
the field $F.$ The field $F$ satisfies the equations%
\begin{equation}
dF=0\qquad\underset{\text{\textbf{g}}}{\delta}F=-J.  \label{5}
\end{equation}
Eqs.(\ref{5}) can be written in a general Lorentzian spacetime, taking into
account Eq.(\ref{4bisss}), as%
\begin{equation}
{\mbox{\boldmath$\partial$}}F=J.  \label{7}
\end{equation}
Neglecting radiation reaction, the motion of an arbitrary probe charge of
mass $(m,q,\sigma)$ is given by 
\begin{equation}
m\nabla_{v}v=qv\lrcorner F,  \label{8}
\end{equation}
where $v\equiv$\textbf{g}$(\sigma_{\ast},\;),$ with $\sigma_{\ast}$ the
tangent vector field to the \textit{worldline} $\sigma:\mathbb{R\rightarrow}%
M $ of the charged particle. Eqs.(\ref{7}) and (\ref{8}) are \textit{%
intrinsic}, i.e., they do not depend on any reference frame and/or
coordinates used by observers living on different reference frames. Note
that the concept of observer is \textit{different} from that of a reference
frame. An observer is modelled by an integral line of a reference frame \cite%
{sawu,rosha}. Indeed, a reference frame $\mathbf{e}_{\mathbf{0}}$ can be
viewed as the four-velocity field of a family of test observers whose
worldlines are the integral curves of $\mathbf{e}_{\mathbf{0}},$ each one
can be parametrized by the proper time $\tau_{\mathbf{e}_{0}},$ defined up
to an additive constant on each curve.

Now, Eqs.(\ref{5}) can be derived from the following Lagrangian density%
\begin{equation}
\mathcal{A=}\int\nolimits_{U}F\wedge\star F-A\wedge\star J.  \label{25}
\end{equation}
As it is well known, every Lagrangian density written in terms of
differential forms is \textit{invariant} under arbitrary diffeomorphisms%
\footnote{%
See, e.g., \cite{thirring}.} $\mathtt{h}:M\rightarrow M$. Under this
diffeomorphism the fields, currents and connection transform under the
pullback mapping, i.e., 
\begin{align}
\eta & \mapsto\text{\textbf{g}}^{\prime}=\mathtt{h}^{\ast}\text{\textbf{g}, }%
\quad A\mapsto\mathtt{h}^{\ast}A,\quad F\mapsto\mathtt{h}^{\ast}F,\quad
J\mapsto\mathtt{h}^{\ast}J,  \label{25bis} \\
\nabla & \mapsto\mathtt{h}^{\ast}\nabla,\quad\mathtt{h}^{\ast}\nabla _{%
\mathtt{h}_{\ast}^{-1}\mathbf{V}}\mathtt{h}^{\ast}\mathbf{t}|_{\mathfrak{e}}=%
\mathtt{h}^{\ast}(\nabla_{\mathbf{V}}\mathbf{t)}|_{\mathtt{h}\mathfrak{e}},
\\
\forall\mathfrak{e} & \in M,\quad\mathbf{V\in}\sec TM,\quad\mathbf{t}\in \sec%
\mathbf{T}M,
\end{align}
where $\mathbf{T}M$ denotes the tensor bundle. The models $(M, $\textbf{g}$%
,\nabla,\tau_{\text{\textbf{g}}\ },\uparrow,A,F,J)$ and $(M,$\textbf{g}$%
^{\prime},\mathtt{h}^{\ast}\nabla,\tau_{\text{\textbf{g}}^{\prime}},\uparrow,%
\mathtt{h}^{\ast}A,\mathtt{h}^{\ast}F,\mathtt{h}^{\ast }J)$ are said to be 
\textit{equivalent} in the sense that if Eqs.(\ref{5}) are satisfied with
well defined initial and boundary conditions then $\mathtt{h}^{\ast}F$
satisfy the equations 
\begin{equation}
d\mathtt{h}^{\ast}F=0\text{,}\qquad\underset{\text{\textbf{g}}^{\prime}}{%
\delta}\mathtt{h}^{\ast}F=-\mathtt{h}^{\ast}J  \label{25biss}
\end{equation}
with well defined transformed initial and boundary conditions. However, take
into account that the equivalence is realized via the introduction of
different universe models that are also declared to be equivalent.

The first formul\ae \, in Eq.(\ref{25biss}) is clearly diffeomorphically
invariant since it is a well known result that $d\mathtt{h}^{\ast}=\mathtt{h}%
^{\ast}d$. The second equation is also diffeomorphically invariant because
the pullback mapping can be represented by an invertible dislocated extensor
field \cite{rool} $h^{-1}:\bigwedge T^{\ast}M\rightarrow\bigwedge T^{\ast}M$
such that its exterior power extension satisfies $\underset{-}{h}^{-1}X=%
\mathtt{h}^{\ast}X,$ for any $X\in\sec{\displaystyle\bigwedge}
T^{\ast}M\hookrightarrow\sec\mathcal{C}\ell(M,\mathtt{g})$ and moreover we
can easily show that \cite{fmr} 
\begin{equation}
\underset{\mathtt{\mathbf{g}}^{\prime}}{\star}=\underset{-}{h}^{-1}\underset{%
\mathtt{\mathbf{g}}}{\star}\underset{-}{h},  \label{25bisss}
\end{equation}
where $\underset{\mathtt{\mathbf{g}}}{\star}$ and $\underset{\mathtt{\mathbf{%
g}}^{\prime}}{\star}$ denote the Hodge star operators
 associated with \textbf{g} and \textbf{g}$^{\prime}$. In this way the equation $\underset{\text{\textbf{g}}^{\prime}}{\delta}\mathtt{h}^{\ast}F=-\mathtt{h}^{\ast}J$ implies the equation $\delta F=J.$ Indeed, we have \begin{equation}
\mathtt{h}^{\ast}F=\underset{\mathbf{g}}{\star^{-1}}d\underset{\mathbf{g}}{%
\star}\mathtt{h}^{\ast}F=\underset{-}{h}^{-1}\star d\star F=\underset{-}{h}%
^{-1}\delta F=h^{-1}J.  \label{25X}
\end{equation}

The \textit{active} formulation of the Principle of Relativity \textit{%
implies} that if the set of geometrical objects $(J,F,(m,e,\sigma ))$ living
on Minkowski spacetime $(M,\mbox{\boldmath{$\eta$}},\nabla ,\tau _{%
\mbox{\boldmath{$\eta$}},}\uparrow )$ satisfies Eqs.(\ref{7}) and (\ref{8}),
with physically realizable initial and boundary conditions, then any other
set $(\bar{J},\bar{F},(m,e,\bar{\sigma}))$, with 
\begin{equation}
\bar{F}=l^{\ast }F,\qquad \bar{J}=l^{\ast }J\qquad \bar{v}=l^{\ast }v,
\label{9}
\end{equation}%
where $l$ a Lorentz mapping, $\mathfrak{e}\mapsto l\mathfrak{e}$, and $%
l^{\ast }$ denotes the pullback mapping, will satisfy%
\begin{equation}
{\mbox{\boldmath$\partial$}}\bar{F}=\bar{J}  \label{10}
\end{equation}%
and%
\begin{equation}
m\nabla _{\bar{v}}\bar{v}=e\bar{v}\lrcorner \bar{F},  \label{11}
\end{equation}%
with also physically realizable initial and boundary conditions. It is
trivial to see, e.g., the validity of Eq. (\ref{10}), for indeed since in
this case, \textbf{g }$=l^{\ast }\eta =\eta $ we have that $\underset{%
\mathtt{\eta }}{\star }l^{\ast }=l^{\ast }\underset{\mathtt{\eta }}{\star }$%
. Note moreover that $l$ is conveniently defined in terms of coordinate
transformations by%
\begin{align}
x^{\prime \mu }(l\mathfrak{e})& =x^{\mu }(\mathfrak{e}),  \label{9bis} \\
x^{\mu }(l\mathfrak{e})& =(\mathbf{L}^{-1}\mathbf{)}_{\nu }^{\mu }x^{\nu }(%
\mathfrak{e}),  \label{9biss}
\end{align}%
where $(\mathbf{L}_{\nu }^{\mu })\in \mathrm{SO}_{1,3}^{e}$ is a Lorentz
transformation. Observe that the coordinate functions $x^{\prime \mu }$
satisfy 
\begin{equation}
{\mbox{\boldmath$\partial$}}x^{\prime \mu }=\gamma ^{\prime \mu }=dx^{\prime
\mu }.  \label{11.bis}
\end{equation}%
These coordinate functions are, of course, naturally adapted coordinates in
the Einstein-Lorentz-Poincar\'{e} gauge to the reference frame $\gamma
_{0}^{\prime }$.

Now consider a velocity boost in the $\gamma_{1}$-direction. \ We write%
\footnote{%
Observe that for that boosts, ${(\mathbf{L}^{-1})}^{\dagger }=\mathbf{L}%
^{-1} $, where $\dagger$ means transpose.} (with $\gamma =(1-\mathtt{v}%
^{2})^{-\frac{1}{2}}$) 
\begin{equation}
\mathbf{L=}\left( 
\begin{array}{cccc}
\gamma & -\mathtt{v}\gamma & 0 & 0 \\ 
-\mathtt{v}\gamma & \gamma & 0 & 0 \\ 
0 & 0 & 1 & 0 \\ 
0 & 0 & 0 & 1%
\end{array}
\right) ,  \label{1new}
\end{equation}%
\begin{equation}
\mathbf{L}^{\mathbf{-}1}\mathbf{=}\left( 
\begin{array}{cccc}
\gamma & \mathtt{v}\gamma & 0 & 0 \\ 
\mathtt{v}\gamma & \gamma & 0 & 0 \\ 
0 & 0 & 1 & 0 \\ 
0 & 0 & 0 & 1%
\end{array}
\right) .  \label{2new}
\end{equation}
Consider, moreover, the frames $\gamma_{0}$, $\gamma_{0}^{\prime}$ and $%
\gamma_{0}^{\prime\prime}$, and the orthonormal sets $\{\gamma_{\mu}\}$, $%
\{\gamma_{\mu}^{\prime}\}$ and $\{\gamma_{\mu}^{\prime\prime}\}$, with%
\footnote{%
Note that we also have $\gamma^{\prime\mu}=R\gamma^{\mu}R^{-1}=\mathbf{L}%
_{\nu}^{\mu}\gamma^{\nu}$, $\gamma^{\prime\mu}=dx^{\prime\mu }=\partial
x^{\prime\mu}$, $\gamma^{\mu}=dx^{\mu}=\partial x^{\mu}$, etc.} 
\begin{align}
\gamma_{\mu}^{\prime} & =R\gamma_{\mu}R^{-1}=(\mathbf{L}^{-1})_{\mu}^{%
\alpha}\gamma_{\alpha}\text{, }\qquad\gamma_{\mu}^{\prime\prime}=R^{-1}%
\gamma_{\mu}R=\mathbf{L}_{\mu}^{\alpha}\gamma_{\alpha},  \label{3new} \\
x^{\prime\mu} & =\mathbf{L}_{\nu}^{\mu}x^{\nu},\quad\qquad\qquad\qquad\qquad
x^{\prime\prime\mu}=(\mathbf{L}^{-1}\mathbf{)}_{\nu}^{\mu}x^{\nu}.
\end{align}
We have in details, 
\begin{align}
\gamma_{0}^{\prime} & =\frac{1}{\sqrt{1-\mathtt{v}^{2}}}(\gamma _{0}+\mathtt{%
v}\gamma_{1}),\;\;\gamma_{1}^{\prime}=\frac{1}{\sqrt {1-\mathtt{v}^{2}}}(%
\mathtt{v}\gamma_{0}+\gamma_{1}),\;\;\gamma_{2}^{\prime
}=\gamma_{2},\;\;\gamma_{3}^{\prime}=\gamma_{3},  \label{24} \\
x^{\prime0} & =\frac{1}{\sqrt{1-\mathtt{v}^{2}}}(x^{0}-\mathtt{v}%
x^{0}),\;\;x^{\prime0}=\frac{1}{\sqrt{1-\mathtt{v}^{2}}}(x^{0}-\mathtt{v}%
x^{0}),\;\;x^{\prime2}=x^{2}\text{, }\;\;x^{\prime3}=x^{3}\text{,}
\label{24bis} \\
\gamma_{0}^{\prime\prime} & =\frac{1}{\sqrt{1-\mathtt{v}^{2}}}(\gamma _{0}-%
\mathtt{v}\gamma_{1})\text{, }\;\;\gamma_{1}^{\prime\prime}=\frac {1}{\sqrt{%
1-\mathtt{v}^{2}}}(\mathtt{v}\gamma_{0}-\gamma_{1}),\;\;\gamma
_{2}^{\prime\prime}=\gamma_{2}\text{, }\;\;\gamma_{3}^{\prime\prime}=%
\gamma_{3}, \\
x^{\prime\prime0} & =\frac{1}{\sqrt{1-\mathtt{v}^{2}}}(x^{0}+\mathtt{v}x^{0})%
\text{, }\;\;x^{\prime0}=\frac{1}{\sqrt{1-\mathtt{v}^{2}}}(x^{0}+\mathtt{v}%
x^{0})\text{, }\;\;x^{\prime\prime2}=x^{2}\text{, }\;\;x^{\prime%
\prime3}=x^{3}.
\end{align}
Now, consider a charge at rest at the origin of the $\gamma_{0}$ frame. Its
field is 
\begin{equation}
\left. F\right\vert _{\mathfrak{e}}=\frac{1}{2}F_{\mu\nu}(x(\mathfrak{e}%
))\gamma^{\mu}\wedge\gamma^{\nu},  \label{4new}
\end{equation}
with 
\begin{align}
F_{0i}(x(\mathfrak{e})) & =q\frac{x^{i}(\mathfrak{e})}{\left\vert \mathbf{x}%
(e)\right\vert ^{3}},\qquad F_{ij}(x(\mathfrak{e}))=0,  \notag \\
\left\vert \mathbf{x}(\mathfrak{e})\right\vert & =\sqrt{\left( x^{1}(%
\mathfrak{e})\right) ^{2}+\left( x^{2}(\mathfrak{e})\right) ^{2}+\left(
x^{3}(\mathfrak{e})\right) ^{2}}.  \label{5new}
\end{align}
\bigskip By definition, for any $u,w\in\sec TM$, 
\begin{equation}
\left. \bar{F}\right\vert _{\mathfrak{e}}(\left. u\right\vert _{\mathfrak{e}%
},\left. w\right\vert _{\mathfrak{e}})=\left. F\right\vert _{\mathfrak{e}%
}(\left. {l_{\ast}u}\right\vert _{l\mathfrak{e}},\left. {l_{\ast}w}%
\right\vert _{l\mathfrak{e}}),  \label{6new}
\end{equation}
from where we get%
\begin{equation}
\bar{F}_{\mu\nu}(x(\mathfrak{e}))=\left( \mathbf{L}^{-1}\right) _{\mu
}^{\alpha}\left( \mathbf{L}^{-1}\right) _{\nu}^{\beta}F_{\mu\nu }(x(l%
\mathfrak{e})).  \label{7new}
\end{equation}
The electric and magnetic parts of the pullback fields in the $\gamma_{0}$
frame are 
\begin{align}
\mathbf{\bar{E}(}x(\mathfrak{e})) & =q\left\{ \frac{x^{1}(l\mathfrak{e})}{%
\left\vert \mathbf{x}\right\vert ^{3}},\gamma\frac{x^{2}(l\mathfrak{e})}{%
\left\vert \mathbf{x}\right\vert ^{3}},\gamma\frac{x^{3}(l\mathfrak{e})}{%
\left\vert \mathbf{x}\right\vert ^{3}}\right\} ,  \notag \\
\mathbf{\bar{B}(}x(\mathfrak{e})) & =q\left\{ 0,\gamma\mathtt{v}\frac {%
x^{3}(l\mathfrak{e})}{\left\vert \mathbf{x}\right\vert ^{3}},-\gamma \mathtt{%
v}\frac{x^{2}(l\mathfrak{e})}{\left\vert \mathbf{x}\right\vert ^{3}}\right\}
,  \label{8new}
\end{align}
and using Eq.(\ref{9biss}) we finally have 
\begin{align}
\mathbf{\bar{E}(}x(\mathfrak{e})) & =q\gamma\left\{ \frac{x^{1}(\mathfrak{e}%
)+\mathtt{v}x^{0}(\mathfrak{e})}{\left[ {\Large R}(\mathfrak{e})\right] ^{3}}%
,\frac{x^{2}(\mathfrak{e})}{\left[ {\Large R}(\mathfrak{e})\right] ^{3}},%
\frac{x^{3}(\mathfrak{e})}{\left[ {\Large R}(\mathfrak{e})\right] ^{3}}%
\right\} ,  \notag \\
\mathbf{\bar{B}(}x(\mathfrak{e})) & =\mathbf{v}\times\mathbf{\bar{E}(}x(%
\mathfrak{e})),  \label{9new}
\end{align}
where $\mathbf{v}=\mathbf{(-\mathtt{v}}, 0, 0\mathbf{)}$ and 
\begin{equation}
{\Large R}(\mathfrak{e})=\sqrt{\gamma^{2}\left( x^{1}(\mathfrak{e})+\mathtt{v%
}x^{0}(\mathfrak{e})\right) ^{2}+\left( x^{2}(\mathfrak{e})\right)
^{2}+\left( x^{3}(\mathfrak{e})\right) ^{2}}.  \label{10new}
\end{equation}
Eqs.(\ref{9new}) give the field of a charge $q$ moving in the negative $%
x^{1} $-direction, as can be calculated directly from the Lienard-Wiechert
potential formul\ae .

We can also write for the field $F,$ 
\begin{equation}
\left. F\right\vert _{\mathfrak{e}}=\frac{1}{2}F_{\mu\nu}(x(\mathfrak{e}%
))\gamma^{\mu}\wedge\gamma^{\nu}=\frac{1}{2}F_{\mu\nu}^{\prime}(x^{\prime }(%
\mathfrak{e}))\gamma^{\prime\mu}\wedge\gamma^{\prime\nu}  \label{11new}
\end{equation}
\noindent where 
\begin{equation}
F_{\mu\nu}^{\prime}(x^{\prime}(\mathfrak{e}))=\left( \mathbf{L}^{-1}\right)
_{\mu}^{\alpha}\left( \mathbf{L}^{-1}\right) _{\nu}^{\beta}F_{\mu\nu }(x(%
\mathfrak{e})),  \label{12new}
\end{equation}
and we have for the electric and magnetic fields in the $\gamma_{0}^{\prime}$
frame,%
\begin{align}
\mathbf{E}^{\prime}\mathbf{(}x^{\prime}(\mathfrak{e})) & =q\gamma\left\{ 
\frac{x^{\prime1}(\mathfrak{e})+\mathtt{v}x^{\prime0}(\mathfrak{e})}{\left[ 
{\Large R}^{\prime}(\mathfrak{e})\right] ^{3}},\frac{x^{\prime2}(\mathfrak{e}%
)}{\left[ {\Large R}^{\prime}(\mathfrak{e})\right] ^{3}},\frac{x^{\prime3}(%
\mathfrak{e})}{\left[ {\Large R}^{\prime}(\mathfrak{e})\right] ^{3}}\right\}
,  \notag \\
\mathbf{B}^{\prime}\mathbf{(}x^{\prime}(\mathfrak{e})) & =\mathbf{v}\times%
\mathbf{E}^{\prime}\mathbf{(}x^{\prime}(\mathfrak{e})),  \label{35a}
\end{align}
with%
\begin{equation}
{\Large R}^{\prime}(\mathfrak{e})=\sqrt{\gamma^{2}\left( x^{\prime 1}(%
\mathfrak{e})+\mathbf{v}x^{\prime0}(\mathfrak{e})\right) ^{2}+\left(
x^{\prime2}(\mathfrak{e})\right) ^{2}+\left( x^{\prime3}(\mathfrak{e}%
)\right) ^{2}}.  \label{13bis}
\end{equation}
We see the $\gamma_{0}^{\prime}$ observers perceive (of course, through
measurements) the field $F$ as the field of a charged particle moving with
constant velocity in the negative $x^{\prime1}$-direction, which is
intuitively obvious. Note that $\gamma_{0}$ observers perceive the field $%
\bar{F}$ in the same way that their colleagues at $\gamma_{0}^{\prime}$
realize $F$. Finally the observers (at rest) in the frame $%
\gamma_{0}^{\prime\prime}$ realize the field $\bar{F}$ as the field of a
particle at rest in that frame.

All these results are {classical}\footnote{%
See specially \cite{anderson}.}, although not explained in general with
rigor.

In definitive, the observers at rest in $\gamma_{0}$ can write%
\begin{align}
F & =\mathbf{E+}\gamma_{5}\mathbf{B}  \label{22} \\
\bar{F} & =\mathbf{\bar{E}+}\gamma_{5}\mathbf{\bar{B},}
\end{align}
with $\mathbf{E=}F^{i0}\mathbf{\sigma}_{i}$, $\mathbf{B=}\frac{1}{2}%
\varepsilon^{ijk}F_{jk}\mathbf{\sigma}_{i}$, $\mathbf{\bar{E}=}\bar{F}^{i0}%
\mathbf{\sigma}_{i}$, $\mathbf{\bar{B}=}\frac{1}{2}\varepsilon^{ijk}\bar{F}%
_{jk}\mathbf{\sigma}_{i}$, $\mathbf{\sigma}_{i}=\gamma_{i}\gamma_{0}$ and
the observers at $\gamma_{0}^{\prime}$ can write\footnote{%
Note that $\gamma^{\prime5}=\gamma^{5}$,}%
\begin{equation}
F=\mathbf{E}^{\prime}+\gamma_{5}^{\prime}\mathbf{B}^{\prime}.  \label{23}
\end{equation}
The relations of all these fields are well-defined and have precise physical
meaning.

\section{Active Local Lorentz Rotations of the Electromagnetic Field}

\label{sec4} Action (\ref{25}) is also invariant under local (i.e.,
spacetime point dependent) Lorentz transformations. This statement is
trivial once we use the Clifford bundle formalism. Indeed, taking into
account that 
\begin{equation}
F\wedge \underset{\text{\textbf{g}}}{\star }F=(F\cdot F)\tau _{\text{\textbf{%
g}}},  \label{26}
\end{equation}%
we see that if we perform an \textit{active }Lorentz transformation 
\begin{equation}
F\mapsto \overset{R}{F}=RFR^{-1},  \label{26'}
\end{equation}%
where $R\in \sec \mathrm{Spin}_{1,3}^{e}(M)\hookrightarrow \sec \mathcal{%
C\ell }^{0}\mathcal{(}M,\mathtt{g})$, since $\tau _{\mathtt{g}}=\gamma ^{%
\mathbf{5}}$ which commutes with even sections of the Clifford bundle, we
have 
\begin{equation}
F\wedge \underset{\text{\textbf{g}}}{\star }F=\overset{R}{F}\wedge \underset{%
\text{\textbf{g}}}{\star }\overset{R}{F}.  \label{27}
\end{equation}%
What is the meaning of the field $\overset{R}{F}$? A trivial calculation, as
shown originally by Hestenes \cite{hestenes}, reveals that in the case where 
$R$ is a constant Lorentz transformation in Minkowski spacetime, the
components of $\overset{R}{F}$ in the $\gamma _{0}$ inertial frame field are
the components of $F$ as seen in the $\gamma _{0}^{\prime }$ inertial frame.
But the important question, that is the source of much confusion in the
literature arises: is $\overset{R}{F}$ a solution of Maxwell equations with
a transformed source term $RJR^{-1}$? The answer in the \textit{Clifford
bundle }$\mathcal{C\ell }(M,\mathtt{\eta })$ formalism is in general
negative. Indeed, if 
\begin{equation}
dF=0,\qquad \underset{\mathtt{\eta }}{\delta }F=-J,  \label{28}
\end{equation}%
in general 
\begin{equation}
d(RFR^{-1})\neq 0\text{, }\qquad \underset{\mathtt{\eta }}{\delta }%
(RFR^{-1})\neq RJR^{-1}.  \label{29}
\end{equation}%
\noindent This can be easily seen in the Clifford bundle formalism, since in
general, 
\begin{equation}
{\mbox{\boldmath$\partial$}}(RFR^{-1})\neq R({\mbox{\boldmath$\partial$}}%
F)R^{-1},  \label{30}
\end{equation}%
because, of course, in general, $\gamma ^{\mu }R\neq R\gamma ^{\mu }$. After
recalling the concept of \textit{generalized} gauge covariant derivatives ($%
G$-connections) in the context of Dirac theory we shall investigate if it is
possible in some sense to generalize Maxwell equation in order to have local
Lorentz invariance.

\section{Covariant Derivative in the Clifford Bundle}

\label{sec6} \label{sec5} Let $\mathbf{\{e}_{\mathbf{a}}\mathbf{\}}$, $%
\mathbf{\{e}_{\mathbf{a}}^{\prime}\mathbf{\}\in}\sec\mathbf{P}_{\mathrm{SO}%
_{1,3}^{e}}(M)$ two orthonormal frames and $\{\mathbf{\theta}^{\mathbf{a}%
}\},\{\mathbf{\theta }^{\prime\mathbf{a}}\}\in\sec P_{\mathrm{SO}%
_{1,3}^{e}}(M)$ the respective dual bases satisfying 
\begin{equation}
\mathbf{\theta}^{\mathbf{a}}(\mathbf{e}_{\mathbf{b}})=\delta_{\mathbf{b}}^{%
\mathbf{a}},\;\;\mathbf{\theta}^{\mathbf{a}}\cdot\mathbf{\theta }^{\mathbf{b}%
}=\eta^{\mathbf{ab}},\;\;\mathbf{\theta}^{\prime\mathbf{a} }(\mathbf{e}_{%
\mathbf{b}}^{\prime})=\delta_{\mathbf{b}}^{\mathbf{a}},\;\; \mathbf{\theta}%
^{\prime\mathbf{a}}\cdot\mathbf{\theta}^{\prime\mathbf{b}}=\eta^{\mathbf{ab}%
}.
\end{equation}

Let be $R\in$ $\mathrm{Spin}_{1,3}^{e}(M)\hookrightarrow\sec\mathcal{C}%
\ell^{0}(M,g),$ i.e., $R\tilde{R}=1$ such that 
\begin{equation}
\mathbf{\theta}^{\prime\mathbf{a}}=R\mathbf{\theta}^{\mathbf{a}}R^{-1}.
\end{equation}

It is well-known that the covariant derivative $\nabla_{\mathbf{X}}$ of a
Clifford multiform $A\in\sec\bigwedge T^{\ast}M\hookrightarrow\sec\mathcal{C}%
\ell(M,g)$ in the direction of the vector field $\mathbf{X\in}\sec TM$ in
the \textit{gauge} determined $\mathbf{\{e}_{\mathbf{a}}\mathbf{\}\in P}_{%
\mathrm{SO}_{1,3}^{e}}(M)$ is given by 
\begin{equation}
\nabla_{\mathbf{X}}A=\partial_{\mathbf{X}}(A)+\frac{1}{2}[\mathbf{\omega }_{%
\mathbf{X}},A],
\end{equation}
where $\partial_{\mathbf{X}}$ is the Pfaff derivative of form fields,
defined by 
\begin{equation}
\partial_{\mathbf{X}}A:=\frac{1}{p!}X(A_{\mu_{1}\ldots\mu_{p}})\mathbf{%
\theta }^{\mu_{1}}\wedge\cdots\wedge\mathbf{\theta}^{\mu_{p}}
\end{equation}
\noindent and where $\mathbf{\omega}_{\mathbf{X}}\in\sec\bigwedge
\nolimits^{2}T^{\ast}M\hookrightarrow\sec\mathcal{C}\ell(M,\mathtt{g})$ is a 
$\bigwedge \nolimits^{2}T^{\ast}M$-valued connection calculated at $\mathbf{X%
}$ in the given gauge.

We define 
\begin{equation}
\nabla_{\mathbf{X}}\mathbf{\theta}^{\mathbf{a}}=\frac{1}{2}[\mathbf{\omega }%
_{\mathbf{X}},\mathbf{\theta}^{\mathbf{a}}], \qquad\nabla_{\mathbf{X}}%
\mathbf{\theta}^{^{\prime}\mathbf{a}}=\frac{1}{2}[\mathbf{\omega}_{\mathbf{X}%
}^{\prime},\mathbf{\theta}^{^{\prime}\mathbf{a}}],
\end{equation}
from where we find that: 
\begin{equation}
\mathbf{\omega}_{\mathbf{X}}^{\prime}=R\mathbf{\omega}_{\mathbf{X}%
}R^{-1}+(\nabla_{\mathbf{X}}R)R^{-1}  \label{connection}
\end{equation}

From the fact that 
\begin{align}
\nabla _{\mathbf{e}_{\mathbf{a}}}\mathbf{\theta }^{\mathbf{b}}& =-\omega _{%
\mathbf{ac}}^{\mathbf{b}}\mathbf{\theta }^{\mathbf{c}}=\frac{1}{2}\left[ 
\mathbf{\omega }_{\mathbf{a}},\mathbf{\theta }^{\mathbf{c}}\right] , \\
\nabla _{\mathbf{e}_{\mathbf{a}}^{\prime }}\mathbf{\theta }^{\prime \mathbf{b%
}}& =-\omega _{\mathbf{ac}}^{\prime \mathbf{b}}\mathbf{\theta }^{\prime 
\mathbf{c}}=\frac{1}{2}\left[ \mathbf{\omega }_{\mathbf{a}}^{\prime },%
\mathbf{\theta }^{\prime \mathbf{c}}\right]
\end{align}%
it follows the expressions 
\begin{align}
\mathbf{\omega }_{\mathbf{e}_{\mathbf{a}}}& =-\frac{1}{2}\omega _{\mathbf{a}%
}^{\mathbf{bc}}\mathbf{\theta }_{\mathbf{b}}\wedge \mathbf{\theta }_{\mathbf{%
c}}\in \sec \bigwedge\nolimits^{2}T^{\ast }M\hookrightarrow \sec \mathcal{C}%
\ell (M,\mathtt{g}),\text{ }\omega _{\mathbf{a}}^{\mathbf{bc}}=-\omega _{%
\mathbf{a}}^{\mathbf{cb}} \\
\mathbf{\omega }_{\mathbf{e}_{\mathbf{a}}}^{\prime }& =-\frac{1}{2}\omega _{%
\mathbf{a}}^{\prime \mathbf{bc}}\mathbf{\theta }_{\mathbf{b}}^{\prime
}\wedge \mathbf{\theta }_{\mathbf{c}}^{\prime }\in \sec \bigwedge
{}^{2}T^{\ast }M\hookrightarrow,\text{ }%
\omega _{\mathbf{a}}^{\prime \mathbf{bc}}=-\omega _{\mathbf{a}}^{\prime 
\mathbf{cb}}
\end{align}

\subsection{Covariant Derivative of Spinor Fields}

The covariant derivative of the representative of a Dirac-Hestenes spinor
field is a kind of gauge covariant derivative. Let us explain what we mean
by this wording.

Let $\mathbf{\nabla}_{\mathbf{e}_{\mathbf{a}}}^{s}$ be the spinor covariant
derivative that acting on sections of the left spin-Clifford bundle, i.e.,
on $\mathbf{\chi}\in\sec\mathcal{C}\ell_{\mathrm{Spin}_{1,3}^{e}}(M,\mathtt{g%
})$ \cite{moro2004}. The representative $\mathbf{\nabla}_{\mathbf{e}_{%
\mathbf{a}}}^{(s)}$ acts on the \textit{gauge }representatives of $\mathbf{%
\chi}$ in the Clifford bundle. Consider two spin coframes $%
\Xi,\Xi^{\prime}\in P_{\mathrm{Spin}_{1,3}^{e}}(M)$, such that $s(\Xi)=\{%
\mathbf{\theta }^{\mathbf{a}}\}$ and $s(\Xi^{\prime})=\{\mathbf{\theta}%
^{\prime\mathbf{a}}\}$ where $s:P_{\mathrm{Spin}_{1,3}^{e}}(M)\rightarrow P_{%
\mathrm{SO}_{1,3}^{e}}(M)$ is the fundamental map connecting those bundles 
\cite{moro2004}. Suppose that two different spinor fields $\mathbf{\chi\in}%
\sec\mathcal{C}\ell_{\mathrm{Spin}_{1,3}^{e}}(M,g)$ and $\mathbf{\Phi}\in\sec%
\mathcal{C}\ell_{\mathrm{Spin}_{1,3}^{e}}(M,\mathtt{g})$ have in the spin
frames $\Xi$ and $\Xi^{\prime}$ the same representative $\chi\in\sec\mathcal{C}\ell(M,\mathtt{g})$. Then in each spin frame the representative of the
spin covariant derivative is given by 
\begin{align}
\nabla_{\mathbf{X}}^{(s)}\chi & =\partial_{\mathbf{X}}\chi+\frac{1}{2}%
\mathbf{\omega}_{\mathbf{V}}\chi,  \notag \\
\nabla_{\mathbf{X}}^{\prime(s)}\chi & =\partial_{\mathbf{X}}\chi+\frac{1}{2}%
\mathbf{\omega}_{\mathbf{V}}^{\prime}\chi,  \label{gauge}
\end{align}
where $\mathbf{\omega}_{\mathbf{X}}^{\prime}$ and $\mathbf{\omega}_{\mathbf{X%
}}$ are related as in Eq.(\ref{connection}) and $\mathbf{X\in}\sec TM.$

Now, let $\psi_{\Xi}\in\sec\mathcal{C}\ell(M,\mathtt{g})$ and $\psi
_{\Xi^{\prime}}=\psi_{\Xi_{0}}R^{-1}\in\sec\mathcal{C}\ell(M,\mathtt{g})$ be
the representatives of $\mathbf{\Psi\in}\sec\mathcal{C}\ell_{\mathrm{Spin}%
_{1,3}^{e}}(M,\mathtt{g})$ in two different spin frames $\Xi$ and $\Xi
^{\prime}.$ We have, as it is easy to verify:%
\begin{equation}
\nabla_{\mathbf{X}}^{\prime(s)}\psi_{\Xi^{\prime}}=(\nabla_{\mathbf{X}%
}^{(s)}\psi_{\Xi})R^{-1},  \label{GAUGE}
\end{equation}
which shows that the representative of the spinor covariant derivative in
the Clifford bundle is a kind of gauge covariant derivative.

\begin{remark}
From now on we call $\nabla _{\mathbf{X}}^{(s)}$ simply the spinor
derivative. In each gauge, if $A\in \sec \mathcal{C}\ell (M,\mathtt{g})$ and 
$\psi _{\Xi }\in \sec \mathcal{C}\ell (M,\mathtt{g})$ is the representative
of $\mathbf{\Psi \in }\sec \mathcal{C}\ell _{\mathrm{Spin}_{1,3}^{e}}(M,%
\mathtt{g})$ we have \cite{moro2004} 
\begin{equation}
\nabla _{\mathbf{X}}^{(s)}(A\psi _{\Xi })=\nabla (_{\mathbf{X}}A)\psi _{\Xi
}+A(\nabla _{\mathbf{X}}^{(s)}\psi _{\Xi })  \label{modular derivation}
\end{equation}
\end{remark}

\section{Dirac-Hestenes Equation on Riemann-Cartan Spacetimes}

\label{sec7} Let $(M,\ $\textbf{g}$,\nabla,\tau_{\text{\textbf{g}}\
},\uparrow)$ be a Riemann-Cartan spacetime. The Dirac-Hestenes Lagrangian
written for a representative $\psi\in\sec$ $\mathcal{C}\ell(M,\mathtt{g})$
in a given gauge of a Dirac-Hestenes spinor field $\Psi\in\mathcal{C}\ell_{\mathrm{Spin}_{1,3}^{e}}(M,\mathtt{g})$ read \cite{rorovaz}

\begin{align}
\mathcal{L}(x,\psi,{\mbox{\boldmath$\partial$}}^{(s)}\psi) & =\mathfrak{L}(x%
\text{ },\psi,{\mbox{\boldmath$\partial$}}^{(s)}\psi)dx^{0}\wedge
dx^{1}\wedge dx^{2}\wedge dx^{3}  \notag \\
& =\left[ ({\mbox{\boldmath$\partial$}}^{(s)}\psi\mathbf{\theta}^{\mathbf{0}}%
\mathbf{\theta}^{\mathbf{2}}\mathbf{\theta}^{\mathbf{1}})\cdot\psi-m\psi%
\cdot\psi\right] \sqrt{\left\vert \det\mathtt{\mathbf{g}}\right\vert }%
dx^{0}\wedge dx^{1}\wedge dx^{2}\wedge dx^{3},  \label{dhrc2}
\end{align}
where $\{x^{\mu}\}$ are the coordinate function of a local chart $%
(U,\varphi) $ of the maximal atlas of $M$ and ${\mbox{\boldmath$\partial$}}%
^{(s)}$ the representative of the spin-Dirac operator in the Clifford bundle
is given by: 
\begin{equation}
{\mbox{\boldmath$\partial$}}^{(s)}{\psi:}=\mathbf{\theta}^{\mathbf{a}%
}\nabla_{\mathbf{e}_{\mathbf{a}}}^{(s)}\psi=\mathbf{\theta}^{\mathbf{a}%
}\left( \partial_{\mathbf{e}_{\mathbf{a}}}\psi+\frac{1}{2}\omega_{\mathbf{e}%
_{\mathbf{a}}}\psi\right) ,  \label{dhrc3}
\end{equation}
with $\mathbf{\theta}^{\mathbf{a}}=h_{\mu}^{\mathbf{a}}dx^{\mu}$. The
variational principle used with the Lagrangian density (Eq.(\ref{dhrc2}))
gives after some algebra \cite{rorovaz}

\begin{equation}
{\mbox{\boldmath$\partial$}}^{(s)}\psi\mathbf{\theta}^{\mathbf{2}}\mathbf{%
\theta}^{\mathbf{1}}+\frac{1}{2}T\psi\mathbf{\theta}^{\mathbf{0}}\mathbf{%
\theta}^{\mathbf{2}}\mathbf{\theta}^{\mathbf{1}}-m\psi\mathbf{\theta }^{%
\mathbf{0}}=0,  \label{dhrc17}
\end{equation}
where%
\begin{equation}
T=T_{\mathbf{ab}}^{\mathbf{b}}{\theta}^{\mathbf{a}}.  \label{dhrc18}
\end{equation}
is called the \textit{torsion covector}. Note that in a Lorentzian manifold $%
T=0$ and we obtain the Dirac-Hestenes equation on a Lorentzian manifold. We
observe moreover that the matrix representation of Eq.(\ref{dhrc17})
coincides with an equation first proposed by Hehl and Datta \cite{hehldatta}%
. Eq.(\ref{dhrc17}) is manifestly covariant under a passive gauge
transformation as it is trivial to verify taking into account Eq.(\ref{gauge}%
). We also recall that spinors transforms as scalars under diffeomorphism 
\cite{weinberg} and thus it is easy to verify that the Dirac-Hestenes
equation is invariant under diffeomorphisms.

We observe yet that, if we tried to get the equation of motion related to a
Dirac-Hestenes spinor field on a Riemann-Cartan spacetime, directly from the
equation on Minkowski spacetime by using the principle of minimal coupling,
we would miss the term $\frac{1}{2}T\psi\mathbf{\theta}^{\mathbf{2}}\mathbf{%
\theta}^{\mathbf{1}}$ appearing in Eq.(\ref{dhrc17}). Is this a bad result?
According to \cite{hehldatta} the answer is yes, because there, a supposed
complete theory, where the $\mathbf{\{\theta}^{\mathbf{a}}\mathbf{\}}$ and
the $\{\omega_{\mathbf{e}_{\mathbf{a}}}\}$ are dynamical fields, the spinor
field generates torsion. To put more spice on this issue, let us next
analyze what active Lorentz invariance would imply.

\section{Meaning of Active Lorentz Invariance of the Dirac-Hestenes
Lagrangian}

In the proposed gauge theories of the gravitational field, it is said that
the Lagrangians and the corresponding equations of motion of physical fields
must be invariant under arbitrary \textit{active} local Lorentz rotations.
In this section we briefly investigate how to mathematically implement such
an hypothesis and what is its meaning for the case of a Dirac-Hestenes
spinor field on a Riemann-Cartan spacetime. The Lagrangian we shall
investigate is the one given by Eq.(\ref{dhrc2}), which we now write with
all indices indicating the representative gauge (i.e., spin coframe)

\begin{equation}
\mathfrak{L}(x\text{ },\psi_{\Xi},{\mbox{\boldmath$\partial$}}%
^{(s)}\psi_{\Xi })=\left[ (\mathbf{\theta}^{\mathbf{a}}\nabla_{\mathbf{e}_{%
\mathbf{a}}}^{(s)}{}\psi_{\Xi}\mathbf{\theta}^{\mathbf{0}}\mathbf{\theta}^{%
\mathbf{2}}\mathbf{\theta}^{\mathbf{1}})\cdot\psi_{\Xi}-m\psi_{\Xi}\cdot%
\psi_{\Xi }\right] \sqrt{\left\vert \det\mathtt{\mathbf{g}}\right\vert }. 
\tag{Dirac-Hestenes}
\end{equation}

Observe that the Dirac-Hestenes Lagrangian has been written in a fixed
(passive gauge) individualized by a spin coframe $\Xi$ and we already know
that it is invariant under passive gauge transformations $%
\psi_{\Xi}\mapsto\psi_{\Xi^{\prime}}=\psi_{\Xi}R^{-1}$ ($R\tilde{R}=1$), $%
R\in \sec\mathrm{Spin}_{1,3}^{e}(M)\hookrightarrow\sec\mathcal{C}\ell(M,\mathtt{g})$ once the `connection' $2$-form $\omega_{V}$ transforms as given
in Eq.(\ref{connection}), i.e., 
\begin{equation}
\frac{1}{2}\omega_{V}\mapsto R\frac{1}{2}\omega_{V}R^{-1}+(\mathbf{\nabla}%
_{V}R)R^{-1}.  \label{passive connection}
\end{equation}

Under an active rotation (gauge) transformation the fields transform in new
fields given by 
\begin{align}
\psi_{\Xi} & \mapsto\psi_{\Xi}^{\prime}=R\psi_{\Xi},\text{ }\psi
_{\Xi^{\prime}}^{\prime}=R\psi_{\Xi}R^{-1}  \notag \\
\mathbf{\theta}^{\mathbf{m}} & \mapsto\mathbf{\theta}^{\prime\mathbf{m}}=R%
\mathbf{\theta}^{\mathbf{m}}R^{-1}=\Lambda_{\mathbf{n}}^{\mathbf{m}}\mathbf{%
\theta}^{\mathbf{n}},  \notag \\
\mathbf{e}_{\mathbf{m}} & \mapsto\mathbf{e}_{\mathbf{m}}^{\prime}=(%
\Lambda^{-1})_{\mathbf{m}}^{\mathbf{n}}\mathbf{e}_{\mathbf{n}}.  \label{ari1}
\end{align}

Now, according to the mathematical ideas behind gauge theories, we must
search for a new connection $\nabla ^{\prime s}$ such that the Lagrangian
results invariant. This will be the case if connections$\ \nabla ^{s}$ and $%
\nabla ^{\prime s}$ are representatives of a $G$-connection as introduced in 
\cite{rorovaz}, i.e.\footnote{%
Note that $\nabla ^{s}$ and $\nabla ^{\prime s}$ are connection in the
spin-Clifford bundle of DHSF whereas $\nabla ^{(s)}$ and $\nabla ^{\prime
(s)}$ the effective covariant derivative operators acting on the
representatives of DHSF in the Clifford bundle.},%
\begin{align}
\nabla _{\mathbf{e}_{\mathbf{m}}^{\prime }}^{\prime (s)}(R\psi _{\Xi })&
=R\nabla _{\mathbf{e}_{\mathbf{m}}}^{(s)}{}\psi _{\Xi }, \\
& \text{or}  \notag \\
\nabla _{\mathbf{e}_{\mathbf{n}}}^{\prime (s)}(R\psi _{\Xi })& =\Lambda _{%
\mathbf{n}}^{\mathbf{m}}R\nabla _{\mathbf{e}_{\mathbf{m}}}^{(s)}{}\psi _{\Xi
},  \label{ari2}
\end{align}

Also, taking into account the structure of a representative of a spinor
covariant derivative in the Clifford bundle \ we may verify that in order
for Eq.(\ref{ari2}) to be satisfied we need that the Pfaff derivative
transforms as%
\begin{equation}
\partial _{\mathbf{e}_{n}}\mapsto \partial _{\mathbf{e}_{n}}^{\prime
}=\Lambda _{\mathbf{n}}^{\mathbf{m}}\partial _{\mathbf{e}_{m}},
\end{equation}%
and that the connection transforms as 
\begin{align}
\omega _{\mathbf{e}_{\mathbf{n}}}^{\prime }& =\Lambda _{\mathbf{n}}^{\mathbf{%
m}}\left( R\omega _{\mathbf{e}_{\mathbf{m}}}R^{-1}-2\mathbf{\partial }_{%
\mathbf{e}_{\mathbf{m}}}(R)R^{-1}\right) ,  \notag \\
& \text{or}  \notag \\
\omega _{\mathbf{e}_{\mathbf{m}}^{\prime }}^{\prime }& =R\omega _{\mathbf{e}%
_{\mathbf{m}}}R^{-1}-2\mathbf{\partial }_{\mathbf{e}_{\mathbf{m}}}(R)R^{-1}.
\label{ari3}
\end{align}

Under these conditions we have:%
\begin{align}
& \left[ (\mathbf{\theta}^{\prime\mathbf{a}}\nabla_{\mathbf{e}_{\mathbf{a}%
}^{\prime}}^{\prime(s)}{}\psi_{\Xi^{\prime}}^{\prime}\mathbf{\theta}^{\prime%
\mathbf{0}}\mathbf{\theta}^{\prime\mathbf{2}}\mathbf{\theta}^{\prime\mathbf{1%
}})\cdot\psi_{\Xi^{\prime}}^{\prime}-m\psi_{\Xi^{\prime}}^{\prime}\cdot%
\psi_{\Xi^{\prime}}^{\prime}\right] \sqrt{\left\vert \det\mathtt{\mathbf{g}}%
\right\vert }  \notag \\
& =\left[ (\mathbf{\theta}^{\prime\mathbf{a}}\nabla_{e_{\mathbf{a}}^{\prime
}}^{\prime(s)}{}\psi_{\Xi}^{\prime}\mathbf{\theta}^{\mathbf{0}}\mathbf{%
\theta }^{\mathbf{2}}\mathbf{\theta}^{\mathbf{1}})\cdot\psi_{\Xi}^{\prime}-m%
\psi _{\Xi}^{\prime}\cdot\psi_{\Xi}^{\prime}\right] \sqrt{\left\vert \det%
\mathtt{\mathbf{g}}\right\vert } \\
& =\left[ (\mathbf{\theta}^{\mathbf{a}}\nabla_{e_{\mathbf{a}%
}}^{(s)}{}\psi_{\Xi}\mathbf{\theta}^{\mathbf{0}}\mathbf{\theta}^{\mathbf{2}}%
\mathbf{\theta}^{\mathbf{1}})\cdot\psi_{\Xi}-m\psi_{\Xi}\cdot\psi_{\Xi }%
\right] \sqrt{\left\vert \det\mathtt{\mathbf{g}}\right\vert },  \notag
\end{align}
and we get 
\begin{equation}
\mathfrak{L}(x,\psi_{\Xi^{\prime}}^{\prime},{\mbox{\boldmath$\partial$}}%
^{\prime(s)}\psi_{\Xi^{\prime}}^{\prime})=\mathfrak{L}(x,\psi_{\Xi },{%
\mbox{\boldmath$\partial$}}^{(s)}\psi_{\Xi}).
\end{equation}

Write now, 
\begin{align}
\omega_{\mathbf{e}_{\mathbf{n}}}^{\prime} & =\frac{1}{2}\omega_{\mathbf{m}%
}^{\prime\mathbf{kl}}\mathbf{\theta}_{\mathbf{k}}\wedge\mathbf{\theta }_{%
\mathbf{l}}=\frac{1}{2}\omega_{\mathbf{m}}^{\prime\mathbf{kl}}\mathbf{\theta}%
_{\mathbf{kl}}\in\sec\mathcal{C}\ell(M,\mathtt{g}),  \notag \\
\omega_{\mathbf{e}_{\mathbf{n}}} & =\frac{1}{2}\omega_{\mathbf{m}}^{\mathbf{%
kl}}\mathbf{\theta}_{\mathbf{k}}\wedge\mathbf{\theta}_{\mathbf{l}}=\frac{1}{2%
}\omega_{\mathbf{m}}^{\mathbf{kl}}\mathbf{\theta}_{\mathbf{kl}}\in\sec%
\mathcal{C}\ell(M,\mathtt{g}),  \notag \\
U & =e^{F},\quad F=\frac{1}{2}F^{\mathbf{rs}}\mathbf{\theta}_{\mathbf{rs}%
}\in\sec\mathcal{C}\ell(M,\mathtt{g}).  \label{ari4}
\end{align}
Recall that 
\begin{align}
\omega_{\mathbf{n}}^{\mathbf{rs}} & =\eta^{\mathbf{ra}}\omega_{\mathbf{anb}%
}\eta^{\mathbf{sb}}=\omega_{\mathbf{nb}}^{\mathbf{r}}\eta^{\mathbf{sb}}, 
\notag \\
\omega_{\mathbf{nk}}^{\mathbf{r}} & =\omega_{\mathbf{n}}^{\mathbf{rs}}\eta_{%
\mathbf{sk}}.  \label{ari5}
\end{align}

Then, from Eqs.(\ref{ari3}), (\ref{ari4}) and (\ref{ari5}) we get%
\begin{equation}
\omega_{\mathbf{nk}}^{\prime\mathbf{r}}=\Lambda_{\mathbf{q}}^{\mathbf{b}%
}\omega_{\mathbf{mb}}^{\mathbf{p}}\Lambda_{\mathbf{p}}^{\mathbf{r}}\Lambda_{%
\mathbf{k}}^{\mathbf{m}}-\eta_{\mathbf{sk}}\Lambda_{\mathbf{k}}^{\mathbf{m}%
}\partial_{\mathbf{e}_{\mathbf{m}}}\left( F^{\mathbf{rs}}\right) .
\label{ari6}
\end{equation}

Now, we recall that the components of the torsion tensors $\mathbf{T}$ and $%
\mathbf{T}^{\prime}$ related to the (tensorial) connections $\nabla$ and $%
\nabla^{\prime}$ in the orthonormal basis $\{\mathbf{e}_{\mathbf{r}%
}\otimes\theta^{\mathbf{n}}\wedge\theta^{\mathbf{k}}\}$ are given by 
\begin{align}
T_{\mathbf{nk}}^{\mathbf{r}} & =\omega_{\mathbf{nk}}^{\mathbf{r}}-\omega_{%
\mathbf{kn}}^{\mathbf{r}}-c_{\mathbf{nk}}^{\mathbf{r}},  \notag \\
T_{\mathbf{nk}}^{\prime\mathbf{r}} & =\omega_{\mathbf{nk}}^{\prime \mathbf{r}%
}-\omega_{\mathbf{kn}}^{\prime\mathbf{r}}-c_{\mathbf{nk}}^{\prime\mathbf{r}},
\label{ari7}
\end{align}
where $[\mathbf{e}_{\mathbf{n}},\mathbf{e}_{\mathbf{k}}]=c_{\mathbf{nk}}^{%
\mathbf{r}}\mathbf{e}_{\mathbf{r}}$.

Let us suppose that we start with a torsion free connection $\nabla$. This
means that $c_{\mathbf{nk}}^{\mathbf{r}}=\omega_{\mathbf{nk}}^{\mathbf{r}%
}-\omega_{\mathbf{kn}}^{\mathbf{r}}$. Then 
\begin{equation}
T_{\mathbf{nk}}^{\prime\mathbf{r}}=\Lambda_{\mathbf{n}}^{\mathbf{b}}\Lambda_{%
\mathbf{k}}^{\mathbf{m}}\Lambda_{\mathbf{p}}^{\mathbf{r}}c_{\mathbf{mb}}^{%
\mathbf{p}}-c_{\mathbf{nk}}^{\mathbf{r}}-\partial _{\mathbf{e}_{\mathbf{m}%
}}\left( F^{\mathbf{rs}}\right) \left[ \eta_{\mathbf{sk}}\Lambda_{\mathbf{n}%
}^{\mathbf{m}}-\eta_{\mathbf{sn}}\Lambda_{\mathbf{k}}^{\mathbf{m}}\right] ,
\label{ari8}
\end{equation}
and we see that $\mathbf{T}^{\prime}=0$ only for very particular gauge
transformations.

We then conclude that to suppose the Dirac-Hestenes Lagrangian is invariant
under active rotational gauge transformations implies in an equivalence
between torsion free and non-torsion free connections. Note also that we may
have equivalence between spacetimes with null and non-null curvatures, as it
is easily to verify. It is always emphasized that in a theory where besides $%
\psi $, also the the tetrad fields $\mathbf{\theta }^{\mathbf{a}}$ and the
connection $\omega $ are dynamical variables, the torsion is not zero,
because its source is the spin of the $\psi $ field. Well, this is true in
particular gauges, because as showed above it seems that it is always
possible to find gauges where the torsion is null.

\subsection{The Case of the Local Lorentz Invariance of the Electromagnetic
Field Equations}

If we are prepared to accept as equivalent spacetimes with different
curvatures and torsion tensors then we can modify Maxwell equations in such
a way that they are formally invariant under local Lorentz transformations.
We start with Maxwell theory on a general Riemann-Cartan spacetime $(M,\ $%
\textbf{g}$,\nabla,\tau_{\text{\textbf{g}}\ },\uparrow),$ where we propose
that Maxwell equation is given by 
\begin{equation*}
{\mbox{\boldmath$\partial$}}F=J,
\end{equation*}
where $F\in\sec\bigwedge \nolimits^{2}T^{\ast}M\hookrightarrow\sec\mathcal{C}\ell(M,\mathtt{g})$ and $J\in\sec\bigwedge
\nolimits^{1}T^{\ast}M\hookrightarrow\sec\mathcal{C}\ell(M,\mathtt{g})$ and $%
{\mbox{\boldmath$\partial$}}=\mathbf{\theta}^{\mathbf{a}}\nabla _{\mathbf{e}%
_{\mathbf{a}}}=d-\delta$ is the Dirac operator in a particular (\textit{%
fiducial} gauge) where the spacetime model is a \textit{Lorentzian} one.

Next we propose that $F$ and all $RFR^{-1}$ are gauge equivalent (in
different but equivalent spacetime models $(M,\ $\textbf{g}$,\nabla,\tau _{%
\text{\textbf{g}}\ },\uparrow)$ and $(M,\ $\textbf{g}$,\nabla^{\prime},\tau_{%
\text{\textbf{g}}\ },\uparrow)^{R}$ and that the Dirac operator in $(M,\ $%
\textbf{g}$,\nabla^{\prime},\tau_{\text{\textbf{g}}\ },\uparrow)^{R}$ is $%
\overset{R}{{\mbox{\boldmath$\partial$}}}=\mathbf{\theta}^{\prime \mathbf{a}%
}\nabla_{\mathbf{e}_{a}^{\prime}}^{\prime}$, $\mathbf{\theta }^{\prime%
\mathbf{a}}=R\mathbf{\theta}^{\mathbf{a}}R^{-1}$, $\mathbf{\theta }^{\prime%
\mathbf{a}}(\mathbf{e}_{\mathbf{b}}^{\prime})=\delta_{\mathbf{b}}^{\mathbf{a}%
}$ and $R\in\sec\mathrm{Spin}_{1,3}^{e}(M)\hookrightarrow \sec\mathcal{C}\ell(M,\mathtt{g})$. As we can easily verify with the formulas of the last
section we have 
\begin{equation}
\overset{R}{{\mbox{\boldmath$\partial$}}}\overset{R}{F}=\overset{R}{J},
\label{32}
\end{equation}
and we may say that distinct electromagnetic fields are also classified as
distinct equivalence classes, where $F$ and $\overset{R}{F}$ represent the
same field in different gauges.

Note finally that \textit{formally} we may say that under a change of gauge
model the Dirac operator transforms as 
\begin{equation}
{\mbox{\boldmath$\partial$}}\mathbf{\mapsto}\overset{R}{{\mbox{\boldmath$%
\partial$}}}=R{\mbox{\boldmath$\partial$}}R^{-1}.  \label{31}
\end{equation}
Such an equation has been used by other authors in the past, but there, its
clear mathematical meaning is lacking. In Appendix we present a context in
which an equation like Eq.(\ref{31}) makes its appearance. %
\renewcommand{\thesection}{\Alph{section}} \setcounter{section}{0} %
\renewcommand{\theequation}{\Alph{section}.\arabic{equation}}

\section{Appendix: Representation $(M\mathbf{,V,}$ $\mathbf{\bullet})$ of
Minkowski Spacetime and Maxwell Equations}

In the affine structure $(M\mathbf{,V,}\mathbf{\bullet})$ if an arbitrary
event $\mathfrak{e}_{0}$ in $M$ is fixed, any other event $\mathfrak{e}\in M$
can be represented by a vector $\mathbf{x}(\mathfrak{e})\in\mathbf{V}$. We
write%
\begin{equation}
\mathbf{x(}\mathfrak{e}\mathbf{)=}\;\mathfrak{e}-\mathfrak{e}_{0}.
\label{33}
\end{equation}

Recall that $(\mathbf{V}$, $\mathbf{\bullet})\simeq\mathbb{R}^{1,3}$. With
the identification given by Eq.(\ref{33}), Clifford fields of \textit{%
multivectors} are now represented by mappings 
\begin{equation}
\mathbf{C:x\mapsto C(x)\in}\bigwedge\mathbb{R}^{1,3}\hookrightarrow \mathbb{R%
}_{1,3}\text{.}  \label{34}
\end{equation}
The constant vector fields $\mathbf{e}_{0}=(1,0,0,0)$, $\mathbf{e}%
_{1}=(0,1,0,0),\mathbf{e}_{2}=(0,0,1,0),\newline
\mathbf{e}_{3}=(0,0,0,1)$ define coordinates $\{x^{\mu}\}$ in the
Einstein-Lorentz gauge for a point event $\mathfrak{e}$, i.e., 
\begin{equation}
\mathbf{x(}\mathfrak{e}\mathbf{)=}\;x^{\mu}(\mathfrak{e})\mathbf{e}_{\mu }
\label{35}
\end{equation}

Under an active Lorentz transformation, generated by $\mathbf{R\in }\;%
\mathrm{Spin}_{1,3}^{e}\subset\mathbb{R}_{1,3}^{0}$ the position vector is
mapped as $\mathbf{x\mapsto x}^{\prime}$. The correct interpretation is that 
$\mathbf{x}^{\prime}$ is the position vector of a point event $\mathfrak{e}%
^{\prime}\neq\mathfrak{e}.$ We have 
\begin{equation}
\mathbf{x(}\mathfrak{e}^{\prime}\mathbf{)=}\;\mathfrak{e}^{\prime }-%
\mathfrak{e}_{0}=\mathbf{Rx(}\mathfrak{e}\mathbf{)\tilde{R}.}  \label{36}
\end{equation}

Thus an active rotation is really a diffeomorphism (with a \textit{fixed
point}, namely $\mathfrak{e}_{0}$) in $M$. The action of $\mathbf{R}$ on a
Clifford field, say, an electromagnetic field $\mathbf{F(x(}\mathfrak{e})%
\mathbf{)\in}\bigwedge^{2}\mathbb{R}^{1,3}\hookrightarrow\mathbb{R}_{1,3}$
must be interpreted as a mapping 
\begin{equation}
\mathbf{F(x(}\mathfrak{e}\mathbf{)})\longmapsto\mathbf{F}^{\prime}(\mathbf{x}%
^{\prime}(\mathfrak{e}^{\prime}))=\mathbf{RF(x(}\mathfrak{e}\mathbf{))\tilde{%
R}.}  \label{37}
\end{equation}

From now on we omit the event labels $\mathfrak{e},\mathfrak{e}^{\prime}$
when no confusion results. Consider a Lorentz boost%
\begin{equation}
\mathbf{R=}\exp\left( \frac{\chi}{2}\mathbf{e}_{1}\mathbf{e}_{0}\right)
\label{38}
\end{equation}
with $\tanh\chi=\left\vert \left\vert \vec{v}\right\vert \right\vert $ , $%
\vec{v}=(\mathtt{v},0,0)$. \ We write%
\begin{equation}
\mathbf{x}^{\prime}=\mathbf{Rx\tilde{R}}\;\mathbf{=}\;x^{\mu}\mathbf{Re}%
_{\mu }\mathbf{\tilde{R}=}\;x^{\mu}\mathbf{e}_{\mu}^{\prime}=x^{\prime\mu}%
\mathbf{e}_{\mu},  \label{39}
\end{equation}
>From Eq.(\ref{39}) we see that the coordinates of the event $\mathfrak{e}%
^{\prime}$ with respect to $\{\mathbf{e}_{\mu}^{\prime}\}$ are the same as
the event $\mathfrak{e}$ with respect to $\{\mathbf{e}_{\mu}\}$. Then the
coordinates $\{x^{\mu}\}$ and $\{x^{\prime\mu}\}$ are related by Eq.(\ref%
{24bis}). Under these conditions, if $\mathbf{F}(\mathbf{x})$ is the field
generated at $\mathbf{x}$ by a charge at rest in the $\mathbf{e}_{0}$ frame
at position $\mathbf{\bar{x}}$, then $\mathbf{F}^{\prime}(\mathbf{x}%
^{\prime})=\mathbf{RF(x)\tilde{R}}$ is the field generated by a charge at
rest in the $\mathbf{e}_{0}^{\prime}$ frame at the point $\mathbf{\bar{x}}%
^{\prime }$. A trivial calculation shows that $\mathbf{e}_{0}$ frame
observers perceive (of course through measurements) the field $\mathbf{F}%
^{\prime}(\mathbf{x}^{\prime})$ as the field generated by a charge moving in
the positive $x$-direction with velocity $\vec{v}=(\mathtt{v},0,0)$.

In the formalism used in this section the Dirac operator is represented by
the vector derivative $\mathbf{\partial}_{\mathbf{x}}$, such that 
\begin{equation}
\mathbf{\partial}_{\mathbf{x}}x^{\mu}=\mathbf{e}^{\mu}\text{, }\qquad 
\mathbf{e}^{\mu}\bullet\mathbf{e}_{\nu}=\delta_{\nu}^{\mu}.  \label{40}
\end{equation}
Hestenes \cite{hestenes} claims that any Lorentz transformation $\mathbf{R}$%
\textbf{\ }sends\footnote{%
We note that in \cite{balou}, authors choose a different approach to the
transformations of $\mathbf{\partial}_{\mathbf{x}}$, $\mathbf{F}$, etc.
Analysis of the meaning of the transformed fields in that case is more
difficult, and will be not discussed here.}%
\begin{equation}
\mathbf{\partial}_{\mathbf{x}}\mapsto\mathbf{\partial}_{\mathbf{x}%
^{\prime}}^{\prime}=\mathbf{R\partial}_{\mathbf{x}}\mathbf{\tilde{R}}
\label{conjecture}
\end{equation}
such that 
\begin{equation}
\mathbf{\partial}_{\mathbf{x}^{\prime}}^{\prime}x^{\mu}=\mathbf{e}%
^{\prime\mu },\qquad\frac{\partial}{\partial x^{\mu}}=\mathbf{e}%
_{\mu}^{\prime}\bullet\mathbf{\partial}_{\mathbf{x}^{\prime}}^{\prime}.
\label{41}
\end{equation}
Then, if $\mathbf{\partial}_{\mathbf{x}}\mathbf{F(x)=J(x})$, we have 
\begin{equation}
\mathbf{\partial}_{\mathbf{x}^{\prime}}^{\prime}\mathbf{F}^{\prime}(\mathbf{x%
}^{\prime})=\mathbf{J}^{\prime}\mathbf{(x}^{\prime}).  \label{41bis}
\end{equation}

We now know which is the mathematical meaning of the operator $\mathbf{%
\partial}_{\mathbf{x}^{\prime}}^{\prime}$ satisfying Eq.(\ref{conjecture}).
It has been given by our theory.

$\allowbreak$

\end{document}